# Cross-scale interaction between microturbulence and fishbone in fusion plasmas

Yuehao Ma,[1] Bin Zhang,[2] Liutian Gao,[1] Pengfei Liu,[3] Jian Bao,[3] Zhihong Lin,[4] Huishan Cai,[1,*] AhDi Liu,[1] Hailin Zhao,[2] and Tao Zhang[2]

[1] *CAS Key Laboratory of Frontier Physics in Controlled Nuclear Fusion, School of Nuclear Science and Technology, University of Science and Technology of China, Hefei 230031, China*

[2] *Institute of Plasma Physics, Chinese Academy of Sciences, Hefei 230031, China*

[3] *Institute of Physics, Chinese Academy of Sciences, Beijing 100190, China*

[4] *Department of Physics and Astronomy, University of California, Irvine, California 92697, USA*

Global gyrokinetic simulations are performed for the first time to investigate cross-scale interactions between electromagnetic ion temperature gradient (ITG) turbulence and fishbone instability in tokamak plasmas. The investigation of fluctuation response in the multiscale simulation including both instabilities indicates a strong impact of fishbone on ITG turbulence. Detailed analysis reveals that fishbone-driven zonal radial electric fields at nonlinear saturation significantly suppress electromagnetic ITG turbulence, reducing ion thermal transport close to the neoclassical level. The simulation results agree well with experimental observations of turbulence suppression during fishbone bursts. These findings advance understanding of multiscale interactions that enhance thermal confinement in fusion plasmas.

## I. INTRODUCTION

Multiscale interactions involving complex nonlinear physical processes are ubiquitous in plasmas, such as magnetic reconnection in space plasmas [1], turbulence driven solar-wind acceleration [2], and the interplay between microturbulence and energetic particles (EPs) in magnetic fusion plasmas [3]. It is well established that microturbulence dominates heat and particle transport in tokamaks, severely constraining plasma confinement [4–7]. In particular, ion temperature gradient (ITG) turbulence dominates core ion heat transport and its control is important for achieving high-confinement operation in the International thermonuclear experimental reactor (ITER) [6–9]. Energetic particles from fusion reactions and auxiliary heating are essential for achieving a self-sustained burning plasma because they provide the core plasma heating [3]. Recent experiments have found that EPs affect thermal plasma confinement, indicating strong interactions with microturbulence [10–14]. The effects of EPs arise not only from their direct impact on ITG turbulence [10,11] but also from interactions between EP-driven unstable modes and ITG turbulence [12–14]. However, understanding of multiscale interactions between ITG turbulence and EP-driven instabilities remains limited, owing to the far broader spatiotemporal scale separation that makes the problem particularly challenging.

Energetic particles can excite mesoscale (energetic-ion gyroradius) Alfvén eigenmodes (AEs) and energetic particle modes (e.g., marcoscopic fishbone instability), driving substantial EP transport that degrades core confinement [15]. Microscale ITG turbulence has characteristic wavelengths of order the thermal-ion gyroradius and frequencies typically much lower than those of EP-driven modes. Despite the separation in spatial and temporal scales, strong cross-scale coupling can arise between them. For example, ITG can affect saturation amplitude of AEs and the associated EP transport in nonlinear interactions [16–18], while AE-driven zonal structures can regulate ITG turbulence [14,19,20]. Beyond mesoscale AEs, the macroscopic fishbone instability [21–23], a notable concern in ITER hybrid scenarios, has been found to facilitate the internal transport barrier (ITB) formation in ASDEX Upgrade [24,25], HL-2A [26,27], and EAST [28–30]. In tokamak plasmas, ITB formation is generally attributed to the suppression of turbulence [5,6,31]. However, whether the fishbone instability regulates turbulent transport is still debated, and the underlying mechanism remains unclear. A hypothesis is that fishbone-driven zonal flows may affect turbulence [32,33]. To elucidate this, a cross-scale simulation of the interaction between ITG turbulence and fishbone is required, but such simulation has never been performed due to significant challenges in multiscale kinetic simulations treating all physical processes on the same footing. Moreover, capturing macroscopic fishbone dynamics requires global simulations and physical treatment of fluctuations near the magnetic axis.

In this work, we report the first cross-scale simulations of both electromagnetic ITG turbulence and fishbone instability using the global gyrokinetic toroidal code (GTC) [34], based on EAST discharges [30] in the ITER-like regime. Our results clearly show that fishbone substantially reduces ITG turbulence saturation amplitude. Fishbone-driven zonal flows generate strong radial electric field shear, which suppresses ITG turbulence and reduces the thermal ion heat transport coefficient ($\chi_i\sim 0.2\,\mathrm{m}^2/\mathrm{s}$) close to the neoclassical level. Moreover, zonal flows play a mediating role in the cross-scale interaction between the fishbone instability and ITG turbulence, thereby regulating turbulent transport. This paradigm may also extend to space and astrophysical plasmas, where zonal-band structures are ubiquitous in planetary atmospheres [35,36].

[*] Contact author: hscai@mail.ustc.edu.cn

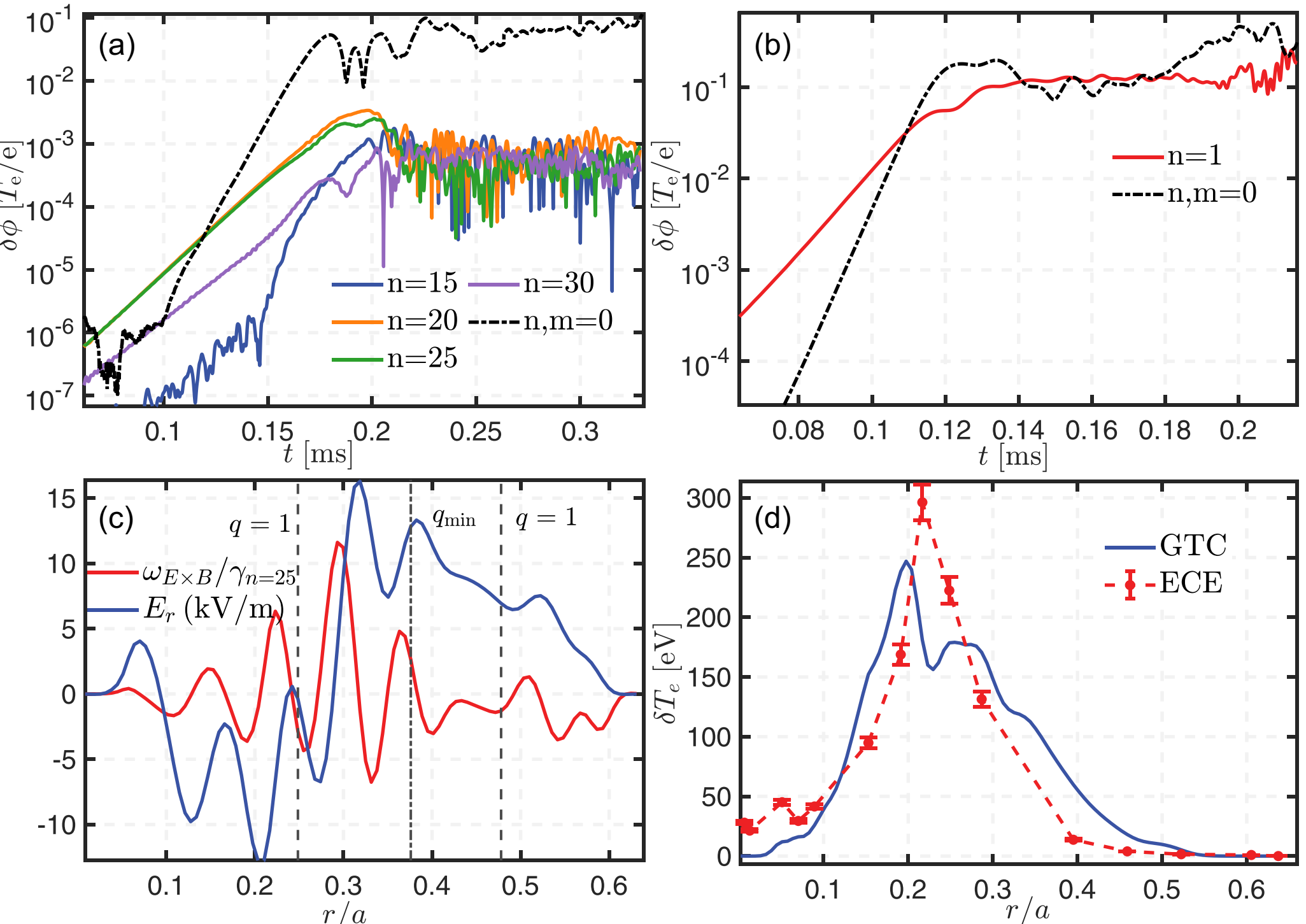

FIG. 1. (a) Time evolution of the perturbed electrostatic potential $\delta\phi$ for selected toroidal modes from ITG-only nonlinear simulations. The black dashed lines denote the zonal potential $\delta\phi_{00}$ (where $n=m=0$, and $m$ is the poloidal mode number), shown as root-mean-square (rms) values across the radial simulation domain. (b) Time evolution of $\delta\phi$ from fishbone-only gyrokinetic simulations. (c) Fishbone-driven radial electric field (blue line) and its $\boldsymbol{E}\times\boldsymbol{B}$ shearing rate (red line) in the nonlinear phase ($t=0.19$ ms), with vertical lines marking the $q=1$ and $q_{\min}$ rational surfaces. The shearing rate is normalized by the linear growth rate of the $n=25$ ITG mode. (d) Electron temperature perturbation $\delta T_e$ at $t=0.19$ ms in fishbone-only simulations, compared with electron cyclotron emission (ECE) measurements.

## II. SIMULATION SETUPS

EAST has conducted a series of low $q_{95}$ discharges targeting ITER hybrid scenarios, where the formation of ion-channel ITB is strongly correlated with the excitation of fishbone [28–30]. Here, we focus on the representative discharge #93890 with $q_{95}\approx 3.4$ [30], and use global gyrokinetic simulations with GTC to investigate the characteristics of turbulence and fishbone, as well as their mutual interaction. The equilibrium geometry and kinetic equilibrium fitting plasma profiles are taken from this discharge at 5000 ms [30]. The safety factor $q$ profile features a weakly reversed magnetic shear, with two $q=1$ surfaces located at $r/a=0.25$ and $r/a=0.48$, and reaches a minimum value of $q_{\min}=0.96$ with $r/a=0.37$, where $r/a$ denotes the normalized minor radius. The major radius is $R_0=1.91$ m, and the minor radius is $a=0.41$ m. In this discharge, the on-axis ion $\beta_i$ reaches 2.02% and the total plasma $\beta_t$ is 5.28%, where $\beta$ denotes the ratio of plasma pressure to magnetic pressure. The energetic ion density profile is obtained from the ONETWO transport code via the NUBEAM module [30], with a neutral beam injection (NBI) energy of approximately 50 keV. On the magnetic axis, the energetic-ion beta fraction and density ratio are $\beta_f/\beta_t=20\%$ and $n_f/n_e=8.5\%$, respectively. In the simulation setup, all plasma species are described by the gyrokinetic framework [37]. We employed a slowing-down distribution for the EPs in simulation. The radial simulation domain extends from $r/a=0$ to $r/a=0.63$. Physical constraints and Fourier series are employed on the perturbation [38], resolving singularity issues associated with the magnetic axis. Numerical convergence studies are meticulously achieved in the GTC simulations. The simulation parameters included radial grids with $N_\psi=100$, poloidal grid points $N_\theta=600$, and a field-aligned mesh consisting of $N_\parallel=32$ parallel grids. A time step size of $1.2\times10^{-8}$ s is selected to ensure the precise resolution of the higher-frequency fishbone. Furthermore, each of the three particle species in the nonlinear electromagnetic simulations was utilized with 500 marker particles per cell.

## III. THE CHARACTERISTICS OF ITG TURBULENCE AND FISHBONE

Global gyrokinetic electrostatic and electromagnetic simulations reveal that the dominant drift-wave instability in the ITB region is the ITG mode [39,40]. Electromagnetic nonlinear simulations (multiple toroidal modes $n=10,11,12,\ldots,35$) are performed to investigate the saturation characteristics of the ITG turbulence. The ITG mode grows near the $q_{\min}$ rational surface, with the most unstable toroidal mode $n = 25$ exhibiting a growth rate $\gamma_{n=25} = 6.8 \times 10^4$ rad/s and a real frequency $\omega_{n=25} = 4.3 \times 10^4$ rad/s. The ITG turbulence saturates via the self-generated zonal flows around 0.2 ms, with a turbulence saturation amplitude $\delta\phi_{n=25} \sim 10^{-3}$ and a zonal potential amplitude $\delta\phi_{00} \sim 5 \times 10^{-2}$, as shown in Fig. 1(a). Here, $\delta\phi$ is given in units of electron temperature $T_e$ divided by elementary charge $e$. The mode structures depicted in Figs. 2(a) and 2(c) show that the ITG turbulence initially features ballooning mode structures located near the $q_{\min}$ rational surface. After nonlinear saturation, the turbulent structures become disordered and spread across the radial domain with shorter correlation lengths.

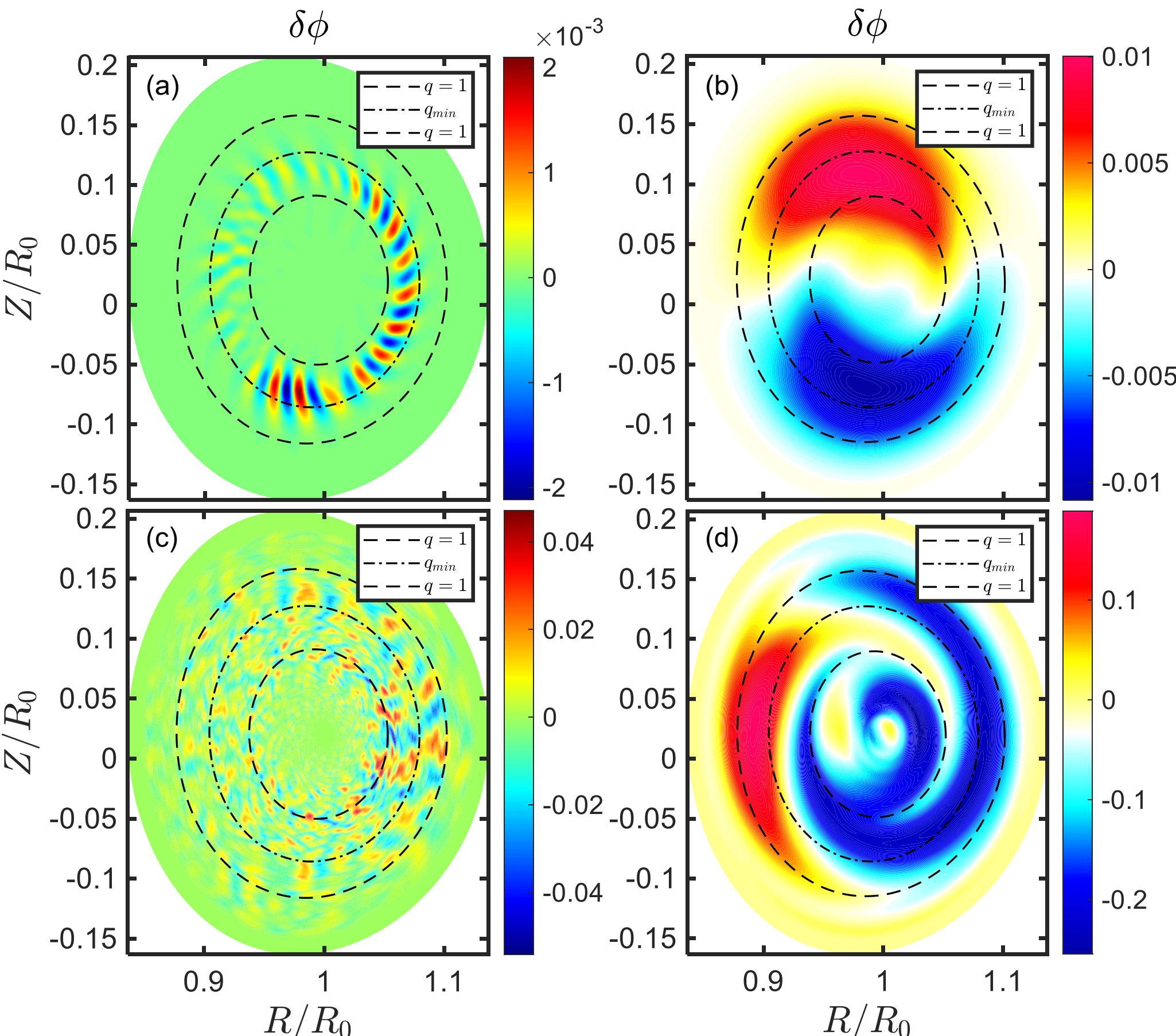

FIG. 2. Poloidal contour plots of electrostatic potential $\delta\phi$ for the ITG-only [panels (a) and (c)] and fishbone-only [panels (b) and (d)] simulations during their linear and nonlinear phases, respectively.

Electromagnetic gyrokinetic simulations including EPs reveal the excitation of fishbone instability. Figure 1(b) shows the time evolution of the perturbed electrostatic potential $\delta\phi$ and zonal potential $\delta\phi_{00}$ in fishbone-only simulation. The dominant $n = m = 1$ fishbone mode exhibits a linear growth rate of $\gamma_{n=1} = 9.4 \times 10^4$ rad/s and mode frequency of $f = 13$ kHz. During the linear and intermediate phases, $\phi_{00}$ grows exponentially at a rate of $\gamma_{n,m=0} \approx 2\,\gamma_{n=1}$, indicating that the zonal flows are generated via a beat-driven process triggered by the fishbone [33,41]. Notably, the zonal flows driven by the fishbone exhibit amplitudes significantly larger than those generated by electromagnetic ITG turbulence. Figure 1(c) shows the fishbone-driven radial electric field and associated $\boldsymbol{E} \times \boldsymbol{B}$ shearing rate $\omega_{E\times B} = -(RB_\theta)^2/B_0\,\partial^2\delta\phi_{00}/\partial\psi^2$ from fishbone-only gyrokinetic simulations [42]. Here, $B_\theta$ is the poloidal magnetic field, $B_0$ is the equilibrium magnetic field, and $\psi$ is the poloidal magnetic flux. This field exhibits a fine-scale radial structure, in contrast to the well-like zonal electric fields obtained in DIII-D simulations [43]. The difference is attributed to the self-consistent inclusion of zonal electron density response in our simulation model. More importantly, the resulting shearing rate $\omega_{E\times B}$ significantly exceeds the linear growth rate of ITG, indicating that radial electric field shear will have a substantial impact on the ITG turbulence.

The fishbone mode structure exhibits macroscopic scales predominantly localized between the two $q = 1$ rational surfaces, as shown in Fig. 2(b). In the nonlinear phase, the mode structure becomes irregular and the radial scale of the modes decreases in Fig. 2(d). The spatial overlap between the turbulence and fishbone in Figs. 2(c) and 2(d) indicates that strong nonlinear interactions may occur. Finally, the electron temperature perturbation envelope $\delta T_e$ obtained from GTC nonlinear simulations agrees with the electron cyclotron emission (ECE) measurements [44,45] in Fig. 1(d). This agreement provides critical experimental validation of the fishbone gyrokinetic simulation results.

## IV. CROSS-SCALE INTERACTIONS OF ITG FISHBONE

To explore cross-scale interactions and accurately evaluate turbulent transport in the ITB region, we performed global nonlinear gyrokinetic simulations that couple electromagnetic ITG turbulence and fishbone. Figure 3(a) shows that in the coupled simulations the fishbone mode saturates first owing to its higher linear growth rate. The zonal flows are primarily driven by the fishbone mode ($\gamma_{n,m=0} \approx 2\gamma_{n=1}$), but their saturation amplitude is slightly lower than in the fishbone-only simulation. The reduction is likely due to partial cancellation between zonal flows generated by intrinsic ITG turbulence and those driven by the fishbone. As shown in Figs. 3(a) and 1(a), the ITG turbulence amplitude $\delta\phi_{n=25}$ is dramatically reduced from $10^{-3}$ to $10^{-4}$ when coupled with fishbone. Correspondingly, Fig. 3(b) shows a dramatic reduction of ITG-induced thermal ion heat conductvity in the presence of fishbone. With fishbone, $\chi_i$ reduces to $\sim 0.2\,\mathrm{m}^2/\mathrm{s}$ within the ITB region near the neoclassical level of ion heat conductvity [30]. The GTC prediction closely matches the ion heat conductvity ($\chi_i \sim 0.3\,\mathrm{m}^2/\mathrm{s}$) inferred from ONETWO power-balance analysis [30]. This turbulence suppression by the fishbone can be confirmed by the mode structure as shown in Fig. 4(a), where the ITG electrostatic potential perturbation manifests as twisted, elongated streamers on very small radial scales. On the other hand, the fishbone is nearly unaffected by ITG turbulence: Both the linear growth rate and the saturation amplitude remain essentially unchanged, as seen by comparing $\delta\phi_{n=1}$ in Figs. 3(a) and 1(b). This is attributed to the fact that ITG-driven fluctuations ($\delta\phi_{n=25} \sim 10^{-4}$) are substantially weaker than those of the fishbone mode ($\delta\phi_{n=1} \sim 10^{-1}$).

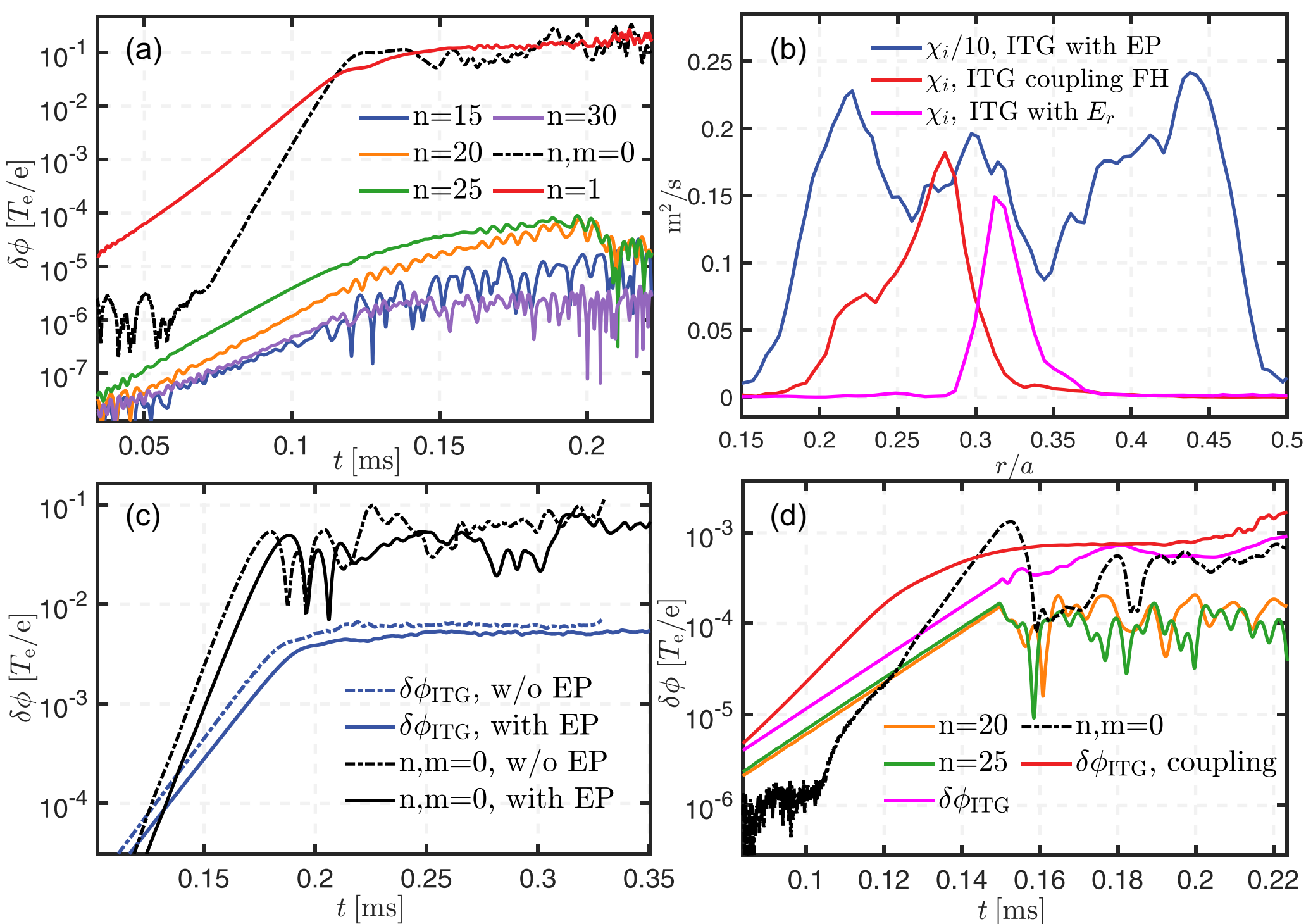

FIG. 3. (a) Time evolution of the $\delta\phi$ in cross-scale coupling simulations. (b) Ion heat conductivity for three cases: ITG simulation with EPs (blue line), cross-scale coupling the fishbone (red line), and ITG-EPs simulation with the fishbone-driven radial electric field $E_r$ (pink line). (c) Time evolution of the volume-averaged turbulence intensity $\delta\phi_{\mathrm{ITG}} = \sqrt{\sum_n (\delta\phi_n)^2}$ and zonal potential from electromagnetic ITG simulations with (solid lines) and without (dashed lines) EPs. (d) Time evolution of $\delta\phi_n$ and turbulence intensity $\delta\phi_{\mathrm{ITG}}$ (pink line) from and ITG-EPs simulation with the fishbone-driven $E_r$. The red line shows the turbulence intensity from the cross-scale coupling simulations.

To understand the suppression mechanism of ITG turbulence, simulations including only EP effects (excluding fishbone by removing the $n = 1$ mode) were first performed. As shown in Fig. 3(c), EPs slightly reduce both the ITG linear growth rate and associated zonal flows, thereby providing weak stabilization to the turbulence. Consequently, turbulence intensity remains comparable to the baseline case without EPs. To clarify the role of fishbone-driven zonal flows, we performed additional electromagnetic ITG simulations that incorporated the radial electric field $E_r$ [from Fig. 1(c)] extracted from fishbone-only simulations. When imposed at the onset of linear phase of ITG evolution ($t \approx 0.15\,\mathrm{ms}$), this $E_r$ shear reduces $\delta\phi_{n=25}$ to $\sim 10^{-4}$, as shown in Fig. 3(d). The resulting suppression of turbulence intensity is consistent with that observed in the coupled cross-scale simulation [compare the pink and red curves in Fig. 3(d)]. Concurrently, the intrinsic zonal flows driven by ITG turbulence is reduced by at least 1 order of magnitude compared to the ITG-only simulation. Notably, the zonal flows are slightly smaller than the turbulence intensity because the fishbone-driven $\boldsymbol{E} \times \boldsymbol{B}$ shearing effect dominates turbulence quenching through eddy decorrelation; thus, the $\delta\phi$ exhibits a small radial correlation length as shown in Fig. 4(b). Finally, we return to Fig. 3(b) to compare the ion heat conductvity $\chi_i$ for the three electromagnetic gyrokinetic cases. Electromagnetic ITG simulation with EPs but without fishbone yields $\chi_i \sim 1.5\,\mathrm{m}^2/\mathrm{s}$ (bule line). Imposing a fishbone-driven $E_r$ profile in the electromagnetic ITG simulation with EPs reduces transport to $\chi_i \sim 0.15\,\mathrm{m}^2/\mathrm{s}$, which is close to the value obtained in the coupled cross-scale simulation. Thus, zonal flows nonlinearly driven by the fishbone provide the dominant coupling channel between macroscale fishbone dynamics and ITG microturbulence, thereby regulating the turbulent transport in hybrid scenarios.

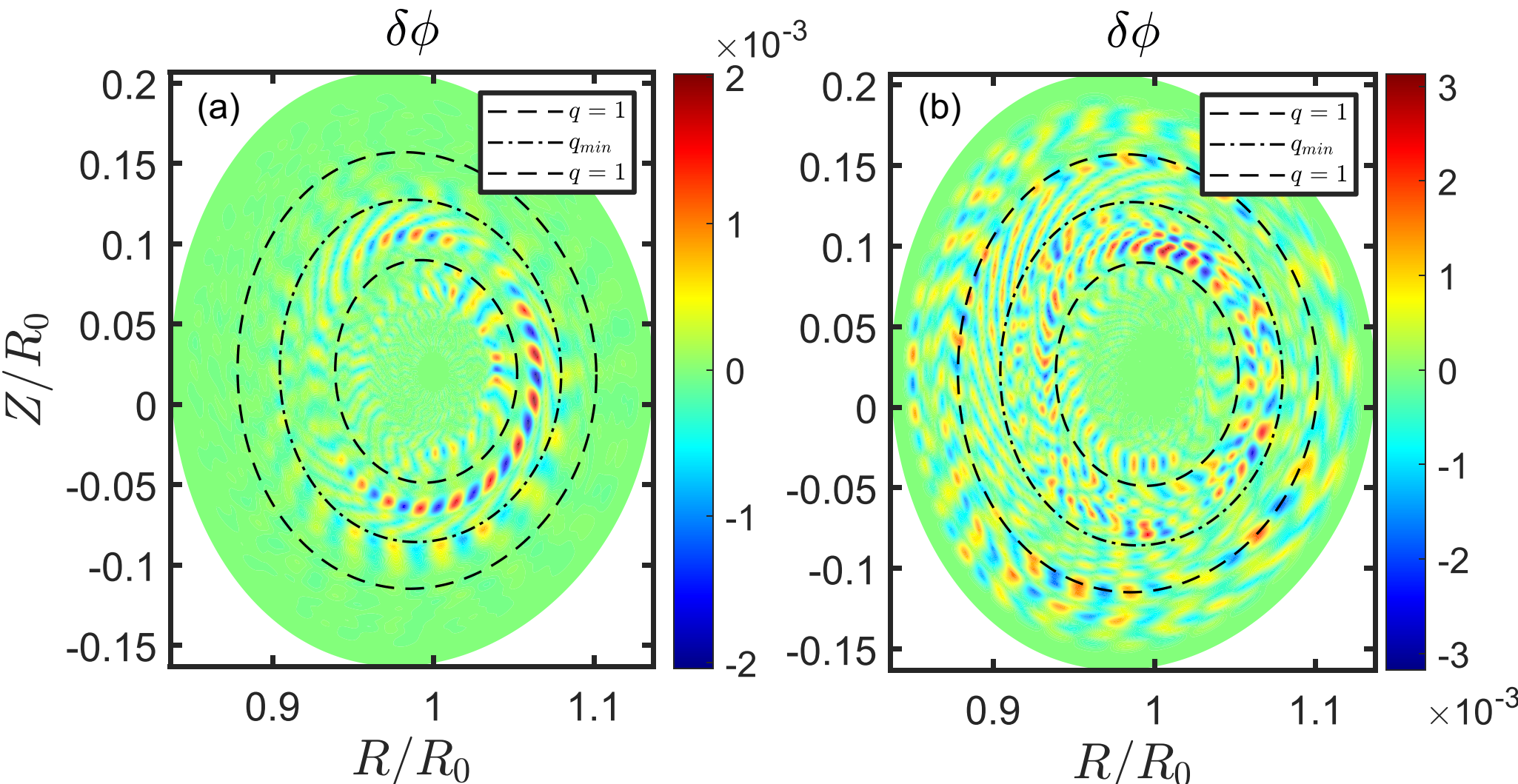

FIG. 4. Poloidal contour plots of $\delta\phi$ for the cross-scale coupling [panel (a)] and ITG-EPs simulations with the fishbone-driven $E_r$ [panel (b)] during their nonlinear phases, respectively.

Experimental evidence from EAST discharge #138189 (Fig. 5) shows that fishbone-driven zonal flows strongly suppress core turbulence. This discharge is similar to #93890, both exhibiting dominant NBI heating and the safety factor profile with weakly reversed magnetic shear. NBI commenced at $t \sim 4.0$ s, and a burst of fishbone activity was subsequently detected by the soft x-ray and Mirnov coils diagnostics around 4.4 s. Concurrently, Doppler reflectometry (DR) measures a pronounced variation in the radial electric field $E_r$ [Fig. 5(b), black], which indicates the onset of fishbone-driven zonal flows [46,47]. The DR probing location is determined by ray tracing to be at $r/a \simeq 0.29$, and the period of this signal coincides with the fishbone bursts. Notably, each successive fishbone burst is correlated with a marked reduction in the turbulence intensity as indicated by a rapid drop in the DR backscattered power. The continuous reduction in turbulence intensity during five fishbone cycles (over 0.4 s) suggests the shearing effect of the zonal flows.

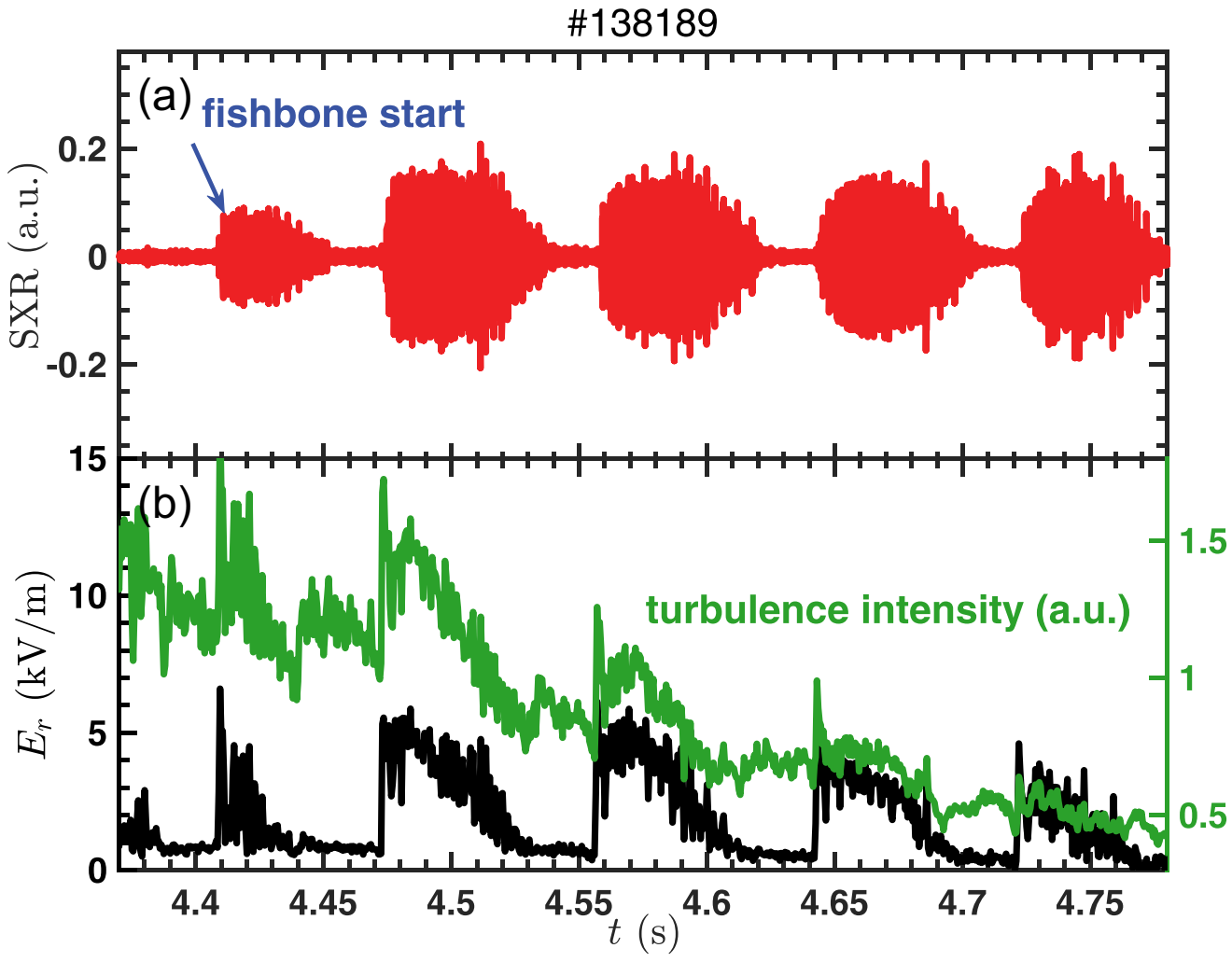

FIG. 5. Experimental evidence from EAST discharge #138189. (a) Time trace of the core-channel soft x-ray (SXR) signal showing fishbone activity (bandpass filtered over the fishbone frequency range). (b) Time traces of DR signals during fishbone activity: The radial electric field $E_r$ inferred from the Doppler shift extracted from the DR phase (black), and the turbulence intensity (green) estimated from the backscattered power.

In summary, GTC simulations and experimental results from EAST demonstrate that fishbone can improve core confinement, underscoring the importance of multiscale interactions. Zonal flows play a mediating role in the cross-scale interaction between fishbone and ITG turbulence. It is of interest that fishbone-driven zonal flows may partially counteract ITG self-generated zonal flows. Thus, careful experimental observations are essential to accurately evaluate the net impact of fishbone on confinement and transport. Nonetheless, these encouraging findings indicate that there is an unexpectedly favorable window for improving plasma confinement in ITER and future fusion reactors. As fishbone in the plasma core can be driven by fusion-born alpha particles, it can strongly affect turbulent transport and plasma performance. Fishbone simultaneously causes redistribution and EP transport, so whether plasma confinement is enhanced or not depends on the balance between these competing effects. The overall impact hinges on equilibrium profiles. Further studies in ITER are underway and will be essential to determine how effectively this mechanism manifests in a burning plasma. The cross-scale interaction between energetic particles and turbulence may also be of interest in space and astrophysical plasmas, such as cosmic-ray origin and transport [48], pickup-ion-driven instabilities in the heliosheath [49], and turbulence-driven solar-wind acceleration [50].

## ACKNOWLEDGMENTS

This work was supported by the National MCF Energy R & D Program of China (Grant No. 2024YFE03050002),

the National Natural Science Foundation of China (Grant No. 12525511), the Strategic Priority Research Program of the Chinese Academy of Sciences (Grants No. XDB0500302 and No. XDB0790202), the Youth Innovation Promotion Association CAS (No. 2023470), and the National Natural Science Foundation of China under Contract No. 12375230. We also acknowledge the support of the Xiyuan Fusion Innovation Fund provided by Fusion Energy Technology Co., Ltd.

## DATA AVAILABILITY

There are no publicly available research data or software supporting this manuscript. Requests for further information or data should be sent to the authors.

[1] D. Biskamp, Magnetic reconnection in plasmas, Astrophys. Space Sci. **242**, 165 (1996).

[2] J. Squire, R. Meyrand, M. W. Kunz, L. Arzamasskiy, A. A. Schekochihin, and E. Quataert, High-frequency heating of the solar wind triggered by low-frequency turbulence, Nat. Astron. **6**, 715 (2022).

[3] M. Salewski *et al.*, Energetic particle physics: Chapter 7 of the special issue: On the path to tokamak burning plasma operation, Nucl. Fusion **65**, 043002 (2025).

[4] W. Horton, Drift waves and transport, Rev. Mod. Phys. **71**, 735 (1999).

[5] J. W. Connor, T. Fukuda, X. Garbet, C. Gormezano, V. Mukhovatov, M. Wakatani, the ITB Database Group, and the ITPA Topical Group on Transport and Internal Barrier Physic, A review of internal transport barrier physics for steady-state operation of tokamaks, Nucl. Fusion **44**, R1 (2004).

[6] K. Ida and T. Fujita, Internal transport barrier in tokamak and helical plasmas, Plasma Phys. Controlled Fusion **60**, 033001 (2018).

[7] M. Yoshida *et al.*, Transport and confinement physics: Chapter 2 of the special issue: On the path to tokamak burning plasma operation, Nucl. Fusion **65**, 033001 (2025).

[8] J. Dong, W. Horton, and J. Kim, Toroidal kinetic $\eta_i$-mode study in high-temperature plasmas, Phys. Fluids B **4**, 1867 (1992).

[9] J. Li and Y. Kishimoto, Interaction between small-scale zonal flows and large-scale turbulence: A theory for ion transport intermittency in tokamak plasmas, Phys. Rev. Lett. **89**, 115002 (2002).

[10] Y.-S. Na, T. Hahm, P. Diamond, A. Di Siena, J. Garcia, and Z. Lin, How fast ions mitigate turbulence and enhance confinement in tokamak fusion plasmas, Nat. Rev. Phys. **7**, 190 (2025).

[11] A. Di Siena *et al.*, New high-confinement regime with fast ions in the core of fusion plasmas, Phys. Rev. Lett. **127**, 025002 (2021).

[12] S. Mazzi *et al.*, Enhanced performance in fusion plasmas through turbulence suppression by megaelectronvolt ions, Nat. Phys. **18**, 776 (2022).

[13] H. Han *et al.*, A sustained high-temperature fusion plasma regime facilitated by fast ions, Nature (London) **609**, 269 (2022).

[14] J. Garcia *et al.*, Stable deuterium-tritium plasmas with improved confinement in the presence of energetic-ion instabilities, Nat. Commun. **15**, 7846 (2024).

[15] L. Chen and F. Zonca, Physics of Alfvén waves and energetic particles in burning plasmas, Rev. Mod. Phys. **88**, 015008 (2016).

[16] W. Heidbrink, J. M. Park, M. Murakami, C. Petty, C. Holcomb, and M. Van Zeeland, Evidence for fast-ion transport by microturbulence, Phys. Rev. Lett. **103**, 175001 (2009).

[17] W. Zhang, Z. Lin, and L. Chen, Transport of energetic particles by microturbulence in magnetized plasmas, Phys. Rev. Lett. **101**, 095001 (2008).

[18] P. Liu, X. Wei, Z. Lin, W. Heidbrink, G. Brochard, G. Choi, J. Nicolau, and W. Zhang, Cross-scale interaction between microturbulence and meso-scale reversed shear Alfvén eigenmodes in DIII-D plasmas, Nucl. Fusion **64**, 076007 (2024).

[19] L. Chen and F. Zonca, Nonlinear excitations of zonal structures by toroidal Alfvén eigenmodes, Phys. Rev. Lett. **109**, 145002 (2012).

[20] X. Du *et al.*, Microturbulence suppression by Alfvén eigenmodes in the DIII-D tokamak, Phys. Rev. Lett. **135**, 265101 (2025).

[21] K. McGuire *et al.*, Study of high-beta magnetohydrodynamic modes and fast-ion losses in PDX, Phys. Rev. Lett. **50**, 891 (1983).

[22] L. Chen, R. White, and M. Rosenbluth, Excitation of internal kink modes by trapped energetic beam ions, Phys. Rev. Lett. **52**, 1122 (1984).

[23] B. Coppi and F. Porcelli, Theoretical model of fishbone oscillations in magnetically confined plasmas, Phys. Rev. Lett. **57**, 2272 (1986).

[24] O. Gruber *et al.*, Stationary $H$-mode discharges with internal transport barrier on ASDEX Upgrade, Phys. Rev. Lett. **83**, 1787 (1999).

[25] S. Günter, A. Gude, J. Hobirk, M. Maraschek, S. Saarelma, S. Schade, R. C. Wolf, and ASDEX Upgrade Team, MHD phenomena in advanced scenarios on ASDEX Upgrade and the influence of localized electron heating and current drive, Nucl. Fusion **41**, 1283 (2001).

[26] W. Chen, Y. Xu, X. T. Ding, Z. B. Shi, M. Jiang, W. L. Zhong, X. Q. Ji, and HL-2A Team, Dynamics between the fishbone instability and nonlocal transient transport in HL-2A NBI plasmas, Nucl. Fusion **56**, 044001 (2016).

[27] W. Deng *et al.*, Investigation of the role of fishbone activity in the formation of internal transport barrier in HL-2A plasma, Phys. Plasmas **29**, 102106 (2022).

[28] X. Gao *et al.*, Sustained high $\beta$ N plasmas on EAST tokamak, Phys. Lett. A **382**, 1242 (2018).

[29] X. Gao *et al.*, Experimental progress of hybrid operational scenario on EAST tokamak, Nucl. Fusion **60**, 102001 (2020).

[30] B. Zhang *et al.*, Progress on physics understanding of improved confinement with fishbone instability at low $q_{95} < 3.5$ operation regime in EAST, Nucl. Fusion **62**, 126064 (2022).

[31] S. Wang, Z. Wang, and T. Wu, Self-organized evolution of the internal transport barrier in ion-temperature-gradient driven gyrokinetic turbulence, Phys. Rev. Lett. **132**, 065106 (2024).

[32] W. Ge, Z.-X. Wang, F. Wang, Z. Liu, and L. Xu, Multiple interactions between fishbone instabilities and internal transport barriers in EAST plasmas, Nucl. Fusion **63**, 016007 (2023).

[33] G. Brochard *et al.*, Saturation of fishbone instability by self-generated zonal flows in tokamak plasmas, Phys. Rev. Lett. **132**, 075101 (2024).

[34] Z. Lin, T. S. Hahm, W. Lee, W. M. Tang, and R. B. White, Turbulent transport reduction by zonal flows: Massively parallel simulations, Science **281**, 1835 (1998).

[35] P. H. Diamond, S. Itoh, K. Itoh, and T. Hahm, Zonal flows in plasma—A review, Plasma Phys. Controlled Fusion **47**, R35 (2005).

[36] T. Nozawa and S. Yoden, Formation of zonal band structure in forced two-dimensional turbulence on a rotating sphere, Phys. Fluids **9**, 2081 (1997).

[37] A. J. Brizard and T. S. Hahm, Foundations of nonlinear gyrokinetic theory, Rev. Mod. Phys. **79**, 421 (2007).

[38] H. R. Lewis and P. M. Bellan, Physical constraints on the coefficients of Fourier expansions in cylindrical coordinates, J. Math. Phys. **31**, 2592 (1990).

[39] Y. Ma, B. Zhang, J. Bao, Z. Lin, W. Zhang, H. Cai, and D. Li, Electrostatic turbulence in EAST plasmas with internal transport barrier, Nucl. Fusion **63**, 056014 (2023).

[40] Y. Ma, P. Liu, J. Bao, Z. Lin, and H. Cai, Electromagnetic turbulence in EAST plasmas with internal transport barrier, Nucl. Fusion **66**, 036007 (2026).

[41] L. Chen, Z. Qiu, and F. Zonca, On beat-driven and spontaneous excitations of zonal flows by drift waves, Phys. Plasmas **31**, 040701 (2024).

[42] T. Hahm, M. Beer, Z. Lin, G. Hammett, W. Lee, and W. Tang, Shearing rate of time-dependent E $\times$ B flow, Phys. Plasmas **6**, 922 (1999).

[43] G. Brochard *et al.*, Saturation of fishbone instability through zonal flows driven by energetic particle transport in tokamak plasmas, Nucl. Fusion **65**, 016052 (2025).

[44] H. Zhao, T. Zhou, Y. Liu, A. Ti, B. Ling, M. E. Austin, S. Houshmandyar, H. Huang, W. L. Rowan, and L. Hu, Upgrade of the ECE diagnostic on EAST, Rev. Sci. Instrum. **89**, 10H111 (2018).

[45] H. Zhao, T. Zhou, Y. Liu, A. Ti, B. Ling, X. Feng, A. Liu, C. Zhou, and L. Hu, A multi-channel correlation ECE system for electron temperature fluctuation measurement on EAST tokamak, Fusion Eng. Des. **149**, 111336 (2019).

[46] L. Gao *et al.*, Upgrades of the W-band Doppler reflectometry for the core region plasma measurement on EAST, Plasma Phys. Controlled Fusion **67**, 055047 (2025).

[47] L. Gao *et al.*, Measurements of the $\vec{E} \times \vec{B}$ velocity fluctuation associated with the fishbone instability using the Doppler reflectometry on EAST, Nucl. Fusion **65**, 116019 (2025).

[48] P. Reichherzer, A. F. Bott, R. J. Ewart, G. Gregori, P. Kempski, M. W. Kunz, and A. A. Schekochihin, Efficient micromirror confinement of sub-teraelectronvolt cosmic rays in galaxy clusters, Nat. Astron. **9**, 438 (2025).

[49] M. Opher, A. Loeb, J. Drake, and G. Toth, A small and round heliosphere suggested by magnetohydrodynamic modelling of pick-up ions, Nat. Astron. **4**, 675 (2020).

[50] T. A. Bowen, T. Ervin, A. Mallet, B. D. Chandran, N. Sioulas, P. A. Isenberg, S. D. Bale, J. Squire, K. G. Klein, and O. Pezzi, Stochastic heating in the sub-Alfvénic solar wind, Phys. Rev. Lett. **135**, 255201 (2025).